\documentclass[preprint,showpacs,showkeys,preprintnumbers,amsmath,amssymb]{revtex4}

\usepackage{graphicx}
\usepackage{dcolumn}
\usepackage{bm}
\usepackage{hyperref}
\hypersetup{colorlinks=false}

\begin{document}


\title{Low-momentum effective interaction in \\ the three-dimensional approach}

\author{S. Bayegan}%
 \email{bayegan@khayam.ut.ac.ir}
 \author{M. Harzchi }
\email{mehdi\underline{ }harzchi@khayam.ut.ac.ir}
\author{M.~R. Hadizadeh}
\email{hadizade@khayam.ut.ac.ir }

\affiliation{ Department of Physics, University of Tehran, P.O.Box
14395-547, Tehran, Iran
}%

\date{\today}

\begin{abstract}
The formulation of the low-momentum effective interaction in the
model space Lee-Suzuki and the renormalization group methods is
implemented in the three-dimensional approach. In this approach
the low-momentum effective interaction $V_{low\,k}$ has been
formulated as a function of the magnitude of momentum vectors and
the angle between them. As an application the spin-isospin
independent Malfliet-Tjon potential has been used into the model
space Lee-Suzuki method and it has been shown that the
low-momentum effective interaction $V_{low\,k}$ reproduces the
same two-body observables obtained by the bare potential $V_{NN}$.

\end{abstract}

\maketitle

\section{Introduction}

One of the most fundamental and essential problems in the nuclear
theory is the derivation of the effective interaction between the
nucleons of nuclei. The conventional nuclear force models are
given by the same one-pion exchange interaction in the long
distance, but differ substantially in their treatment of the short
distance. The fact that the different short distance structures
reproduce the same observables for two-body problem indicates that
low energy observables are insensitive to the details of the short
distance dynamics. This insensitivity is a result of the
separation of the high and low energy scales in the nuclear force,
and implies that we can derive a low-momentum effective
interaction.

During the past years, several methods have been developed to
derive the energy independent low-momentum effective interaction
such as the renormalization group (RG) and the model space
techniques. These approaches are mainly based on a partial wave
(PW) decomposition and the details have been given in references
[1-11]. Bogner \emph{et al}. have developed a low-momentum
effective interaction which is quite successful in describing the
two-nucleon system at low energy. This effective interaction is
independent of the potential models as the cutoff is lowered to
$\Lambda=2.1\,\,fm^{-1}$ \cite{Bogner-PR386,Bogner-PLB576}.

Recently the three-dimensional (3D) approach, which greatly
simplifies the numerical calculations of few-body systems without
performing the PW decomposition, is developed for few-body bound
and scattering problems [12-27]. The motivation for developing
this approach is introducing a direct solution of the integral
equations avoiding the very involved angular momentum algebra
occurring for the permutations, transformations and especially for
the three-body forces. Conceptually the 3D formalism consider all
partial wave channels automatically. Based on this non PW method
the bound and scattering amplitudes are formulated as function of
momentum vector variables, specially their magnitudes and the
angle between them.

In this article our aim is to generate the low-momentum effective
interaction directly in a 3D formalism, where as a simplification
the spin and isospin degrees of freedom have been neglected in the
first step. So the low-momentum effective interaction has been
formulated for spinless identical particles as function of vector
Jacobi momenta. Considering the spin and isospin degrees of
freedom is a major additional task, which will increase more
degrees of freedom into the states \cite{Bayegan-PRC77}. The
formulation of the low-momentum effective interaction in a
spin-isospin dependent 3D approach, based on helicity
representation, is currently underway and it will be reported
elsewhere \cite{Bayegan-p}.

This article is organized as follows. In the section
\ref{sec:Lee-Suzuki method} the model space Lee-Suzuki method has
been used to derive the energy independent model space effective
interaction in the 3D representation. In addition a
renormalization group decimation method has been performed in the
3D representation and a flow equation for the low-momentum
effective interaction $V_{low\,k}$ has been obtained as a function
of momentum vectors in the appendix \ref{appendix:RG method}. The
section \ref{sec:Numerical results} describes the details of the
discussion and numerical calculations of the low-momentum
effective interaction $V_{lowk}$ in the model space Lee-Suzuki
method by using the spin-isospin independent toy Malfliet-Tjon
potential. Finally a summary is given in the section
\ref{sec:summary} and an outlook is provided.

\section{Model space Lee-Suzuki method for $V_{low\,k}$ in the 3D momentum
representation}\label{sec:Lee-Suzuki method}

Several model space techniques such as the Bloch-Horowitz
\cite{Bogner-PR386,Jennings-EL72}, the Unitary Transformation
\cite{Epelbaum-PLB439,Fujii-PRC70} and the Lee-Suzuki [7-11] based
on the restricted space, have been developed for the construction
of the low-momentum effective interaction $V_{low\,k}$.

In this work the Lee-Suzuki formalism has been applied to the free
space nucleon-nucleon problem in the 3D momentum representation
and the low-momentum effective interaction $V_{low\,k}$ has been
obtained as function of the momentum vectors. A momentum-space
Hamiltonian for the full-space two-body problem has been
considered as follows:
\begin{eqnarray}
H(\textbf{k},\textbf{k}')=H_{0}(\textbf{k})\delta(\textbf{k}-\textbf{k}')+V_{NN}(\textbf{k},\textbf{k}'),
\end{eqnarray}
where $H_{0}$ denotes the kinetic energy and
$V_{NN}(\textbf{k},\textbf{k}')$ is the bare two-body interaction.
In the model space methods the projection operators onto the
physically important low-energy model space, the $P$ space, and
the high-energy complement, the $Q$ space, have been introduced
as:
\begin{eqnarray}
\nonumber P &=& \int
d^{3}k\,|\textbf{k}\rangle\langle\textbf{k}|,|\textbf{k}|\leq
\Lambda, \\ Q &=& \int
d^{3}k\,|\textbf{k}\rangle\langle\textbf{k}|,|\textbf{k}|>\Lambda,
\end{eqnarray}
where $\Lambda$ is a momentum cutoff which divides the Hilbert
space into the low and high momentum states. These projections
satisfy the relations:
\begin{eqnarray}
P+Q &=& 1, \nonumber \\  PQ&=&QP=0, \nonumber \\  P^{2}&=&P, \nonumber \\
 Q^{2}&=&Q,
\end{eqnarray}
and they act on the full-space two-body problem states as:
\begin{eqnarray}
\nonumber P|\psi_{\textbf{k}}^{NN}\rangle &=&
|\Psi_{\textbf{k}}^{NN}\rangle,
\\
Q|\psi_{\textbf{k}}^{NN}\rangle &=&
\omega|\Psi_{\textbf{k}}^{NN}\rangle,
\end{eqnarray}
where $|\psi_{\textbf{k}}^{NN}\rangle$ and
$|\Psi_{\textbf{k}}^{NN}\rangle$ denote the states of the full and
model spaces respectively. The $\omega$ is an operator which
transforms the states of the $P$ space to the states of the $Q$
space. The key aspect of the Lee-Suzuki method is the
determination of the $\omega$ operator defined by the following
equation \cite{Jennings-EL72}:
\begin{eqnarray} \label{eq.key-omega}
Q|\psi_{\textbf{k}}^{NN}\rangle=Q\,\omega
P|\psi_{\textbf{k}}^{NN}\rangle.
\end{eqnarray}
The Schr\"{o}dinger equation for the full-space two-body problem
by considering the equation (\ref{eq.key-omega}) can be written
as:
\begin{eqnarray}
H|\psi_{\textbf{k}}^{NN}\rangle=H(P+Q)|\psi_{\textbf{k}}^{NN}\rangle=H(P+Q\,\omega
P)|\psi_{\textbf{k}}^{NN}\rangle=E|\psi_{\textbf{k}}^{NN}\rangle.
\end{eqnarray}
By acting the $P$ operator on the left side of the last equation
the full-space two-body problem has been reduced to the model
space two-body problem of the following form:
\begin{eqnarray}\label{eq.model space SE}
PH(P+Q\,\omega
P)|\psi_{\textbf{k}}^{NN}\rangle=(PH_{0}P+PVP+PVQ\,\omega
P)|\psi_{\textbf{k}}^{NN}\rangle=EP|\psi_{\textbf{k}}^{NN}\rangle,
\end{eqnarray}
where we assumed that the $P$ and $Q$ operators commute with the
$H_{0}$. Therefore the non-hermitian low-momentum effective
potential in the model space that reproduces the model space
components of the wave function from the full-space wave function
can be written as [7-11]:
\begin{eqnarray}
V_{low\,k}=PV(P+Q\omega P).
\end{eqnarray}
By using the integral form of the projection operators $P$ and
$Q$, the low-momentum effective interaction $V_{low\,k}$ can be
written in the 3D representation as:
\begin{eqnarray} \label{eq.Vlowk}
V_{low\,k}(\textbf{k}',\textbf{k})=V_{NN}(\textbf{k}',\textbf{k})+\int_{\Lambda}^{\infty}
dq\,q^{2}\int
d\hat{\textbf{q}}\,\,V_{NN}(\textbf{k}',\textbf{q})\,\omega(\textbf{q},\textbf{k}),
\end{eqnarray}
where \textbf{q} is the momentum vector in the complement model
space $Q$. For an application we can choose the suitable
coordinate system which the vector \textbf{k} is along the $z$
axis and the vector $\textbf{k}'$ is in $x-z$ plane. So the
equation (\ref{eq.Vlowk}) can be rewritten as:
\begin{eqnarray} \label{eq.Vlowk-3D}
\nonumber
V_{low\,k}(k',k,x)=V_{NN}(k',k,x)+\int_{\Lambda}^{\infty}
dq\,q^{2}\int_{-1}^{1}
dx'\int_{0}^{2\pi}d\varphi\,\,V_{NN}(k',q,y)\,\omega(q,k,x'),\\
\end{eqnarray}
where:
\begin{eqnarray}
 x&=&\hat{\textbf{k}} \cdot \hat{\textbf{k}}',\nonumber \\
x'&=&\hat{\textbf{k}}\cdot\hat{\textbf{q}},\nonumber \\
y&=&\hat{\textbf{k}}'\cdot\hat{\textbf{q}}=xx'+\sqrt{1-x^{2}}\sqrt{1-x'^{2}}\cos\varphi,
\nonumber \\
k&=&|\,\textbf{k}|, \nonumber \\ k'&=&|\,\textbf{k}'|, \nonumber \\
q&=&|\,\textbf{q}|.
\end{eqnarray}
To calculate the low-momentum effective potential $V_{low\,k}$ we
need to determine $\omega$. To this aim by applying
$\langle\textbf{q}|$ on the left side of equation
(\ref{eq.key-omega}) and using the integral form of projection
operators $P$ and $Q$, this equation can be rewritten as follows:
\begin{eqnarray} \label{eq.Q component psai}
\Psi_{\textbf{k}}^{NN}(\textbf{q})=\int_{0}^{\Lambda}dpp^{2}\int
d\hat{\textbf{p}}\omega(\textbf{q},\textbf{p})\,\Psi_{\textbf{k}}^{NN}(\textbf{p}),
\end{eqnarray}
where \textbf{p} is the momentum vector in  the model space. We
use the completeness relation in the model space as:
\begin{eqnarray} \label{eq.completeness relation}
\int_{0}^{\Lambda}dk\,k^{2}\int
d\,\hat{\textbf{k}}\,\,\,\tilde{\Psi}_{\textbf{k}}^{NN}(\textbf{p}')\Psi_{\textbf{k}}^{NN}(\textbf{p})
=\delta(\textbf{p}'-\textbf{p}).
\end{eqnarray}
The equation (\ref{eq.Q component psai}) after implementing
equation (\ref{eq.completeness relation}) can be written as:
\begin{eqnarray} \label{eq.omega}
\omega(\textbf{q},\textbf{p})=\int_{0}^{\Lambda}dkk^{2}\int
d\hat{\textbf{k}}\,\,\,\Psi_{\textbf{k}}^{NN}(\textbf{q})\tilde{\Psi}_{\textbf{k}}^{NN}(\textbf{p}),
\end{eqnarray}
where $\Psi_{\textbf{k}}^{NN}(\textbf{p})$ and
$\Psi_{\textbf{k}}^{NN}(\textbf{q})$ are the wave function
components of the $P$ and $Q$ spaces of the full-space
respectively which in the form of the half-on-shell (HOS) two-body
$T$ matrix are given by:
\begin{eqnarray}
\Psi_{\textbf{k}}^{NN}(\textbf{q})&=&
\frac{T(\textbf{q},\textbf{k},k^{2})}{k^{2}-q^{2}},
\end{eqnarray}
\begin{eqnarray}
\Psi_{\textbf{k}}^{NN}(\textbf{p})&=&\delta(\textbf{p}-\textbf{k})+
\frac{T(\textbf{p},\textbf{k},k^{2})}{k^{2}-p^{2}+i\varepsilon}
\nonumber\\&=&
\delta(\textbf{p}-\textbf{k})+\wp\frac{T(\textbf{p},\textbf{k},k^{2})}{k^{2}-p^{2}}
-i\pi\,\delta(k^{2}-p^{2})\,T(\textbf{p},\textbf{k},k^{2}),
\end{eqnarray}
where $\wp$ denotes a principle value. The HOS two-body $T$ matrix
can be obtained from the Lippmann-Schwinger equation in the 3D
representation, which is given as \cite{Elster-FBS24}:
\begin{eqnarray} \label{eq.T-matrix}
T(\textbf{k}',\textbf{k},k^{2})=V_{NN}(\textbf{k}',\textbf{k})+
\int\emph{d}^{3}k''
\frac{V_{NN}(\textbf{k}',\textbf{k}'')\,T(\textbf{k}'',\textbf{k},k^{2})}{k^{2}-k''^{2}+i\varepsilon}.
\end{eqnarray}
In order to solve the equation (\ref{eq.omega}) we have chosen
suitable coordinate system which the vector \textbf{q} is along
the $z$ axis and the vector $\textbf{p}$ is in $x-z$ plane.
Therefore the equation (\ref{eq.omega}) can be rewritten as:
\begin{eqnarray} \label{eq.omega-3D}
\omega(q,p,x)=\int_{0}^{\Lambda}dkk^{2}\int_{-1}^{1}
dx'\int_{0}^{2\pi}d\varphi\Psi_{k}^{NN}(q,x')\tilde{\Psi}_{k}^{NN}(p,y),
\end{eqnarray}
where:
\begin{eqnarray}
\nonumber x&=&\hat{\textbf{q}}\cdot\hat{\textbf{p}},\\ \nonumber
x'&=&\hat{\textbf{q}}\cdot\hat{\textbf{k}},\\ \nonumber
y&=&\hat{\textbf{k}}\cdot\hat{\textbf{p}}=xx'+\sqrt{1-x^{2}}\sqrt{1-x'^{2}}\cos\varphi,\\
\nonumber
k&=&|\,\textbf{k}|,\\ \nonumber p&=&|\,\textbf{p}|,\\
q&=&|\,\textbf{q}|.
\end{eqnarray}
We should mention that the Lee-Suzuki method reproduces the HOS
two-body \emph{T} matrix the same as the RG method, therefore the
solution of this approach has been proven to be equivalent to the
solution of the RG equation \cite{Bogner-nt,Bogner-PLB500}. In
appendix (\ref{appendix:RG method}) we derive a renormalization
group decimation method in the 3D representation and we obtain a
flow equation for the low-momentum effective interaction
$V_{low\,k}$ for future applications.

\section{Discussion and Numerical results }\label{sec:Numerical results}

In our calculations we employ the spin-isospin independent
Malfliet-Tjon potential. This force is a superposition of a
short-range repulsive and long-range attractive Yukawa
interactions. It is given as \cite{Malfliet-NPA127}:
\begin{eqnarray}
V_{NN}(\textbf{k}',\textbf{k})=\frac{1}{2\pi^{2}}\,(\frac{V_{R}}{(\textbf{k}'
-\textbf{k})^{2}+\mu_{R}^{2}}-\frac{V_{A}}{(\textbf{k}'
-\textbf{k})^{2}+\mu_{A}^{2}}).
\end{eqnarray}
With the strengths and the masses of the meson exchange as follow:
$V_{R}=7.291,V_{A}=3.177,\mu_{R}=613.7MeV,\mu_{A}=305.9MeV$. This
potential supports one bound state at $E=-2.23MeV$. With this
interaction we first implement the LU factorization into equation
(\ref{eq.completeness relation}) to calculate
$\tilde{\Psi}_{k}^{NN}(p',x',\varphi')$ as an inverse of the wave
function $\Psi_{k}^{NN}(p,x,\varphi)$ in the model space. Then by
solving the equation (\ref{eq.omega-3D}) we calculate
$\omega(q,p,x)$ and finally we input the $\omega$ operator into
equation (\ref{eq.Vlowk-3D}) to obtain the low-momentum effective
interaction $V_{low\,k}(p',p,x)$.

\begin{figure}
\begin{center}
\includegraphics*[width=10cm]{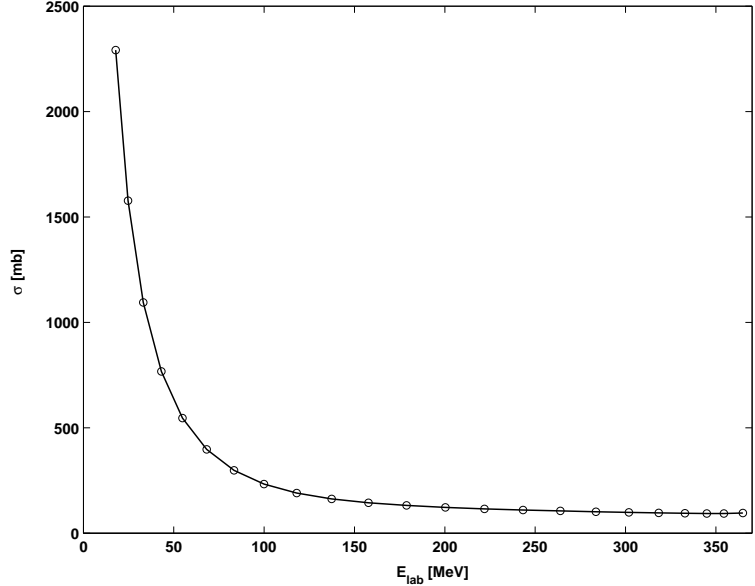}
\caption{\label{fig.cross section}The total two-body cross section
from the low-momentum effective potential (circles) and the bare
potential (solid line) as a function of kinetic energy in the lab
frame.}
\end{center}
\end{figure}

\begin{figure}
\begin{center}
\includegraphics*[width=10cm]{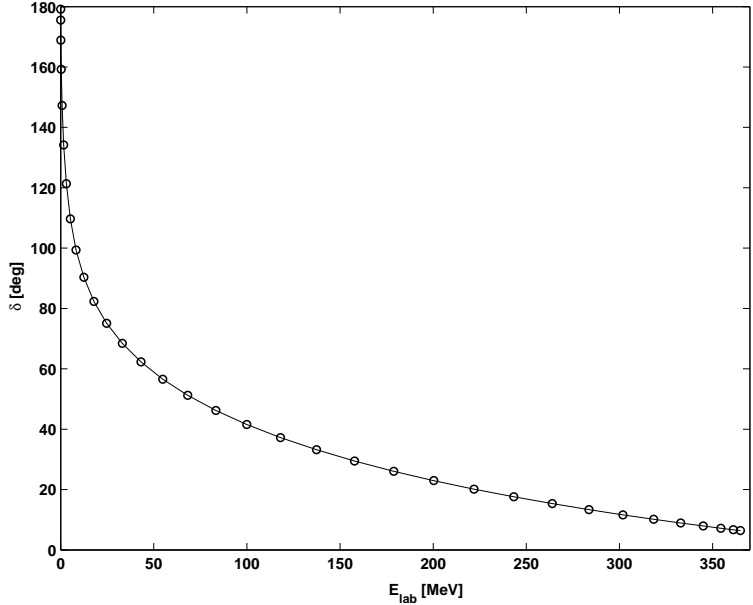}
\caption{\label{fig.phase shift}The two-body $S$-wave phase shift
from the low-momentum effective potential (circles) and the bare
potential (solid line) as a function of kinetic energy in the lab
frame.}
\end{center}
\end{figure}

The dependence on the continuous momentum and the angle variables
is replaced in the numerical treatment by a dependence on the
certain discrete values. To this aim we use the Gaussian
quadrature grid points to discrete the momentum and the angle
variables. In our calculations we choose forty grid points for the
momentum variables in the $P$ space, i.e. the interval
$[0,\Lambda]$, and thirty grid points for the momentum variables
in the $Q$ space, i.e. the interval $[\Lambda,\infty)$. Also
twenty and fourteen grid points for the spherical and the polar
angle variables have been used respectively. The integration
interval for the $P$ and $Q$ spaces are covered by two different
linear and tangential mappings the Gauss-Legendre points $x$ from
the interval (-1,+1) via:
\begin{eqnarray}
p &=& \frac{\Lambda}{2} (1+x),  \nonumber \\ q &=& b \, tan \left(
\frac{\pi}{4} (1+x) \right) + \Lambda \,\, , \, b = 5 \,fm^{-1}.
\end{eqnarray}
to the intervals $[0,\Lambda]$ and $[\Lambda,\infty)$
respectively. As we mentioned in the introduction we have used the
value of $2.1\,fm^{-1}$ for the cutoff $\Lambda$ in our
calculations. The solution of the integral equation
(\ref{eq.T-matrix}) requires a one-dimensional interpolation on
$T$. We use the cubic hermitian splines of ref. \cite{59} for its
accuracy and high computational speed. It can be useful to mention
that in the numerical calculations we use the Lapack library
\cite{60}, for solving a system of linear equations in the
calculation of the two-body $T$ matrix.

\begin{figure}
\begin{center}
\includegraphics*[width=14cm]{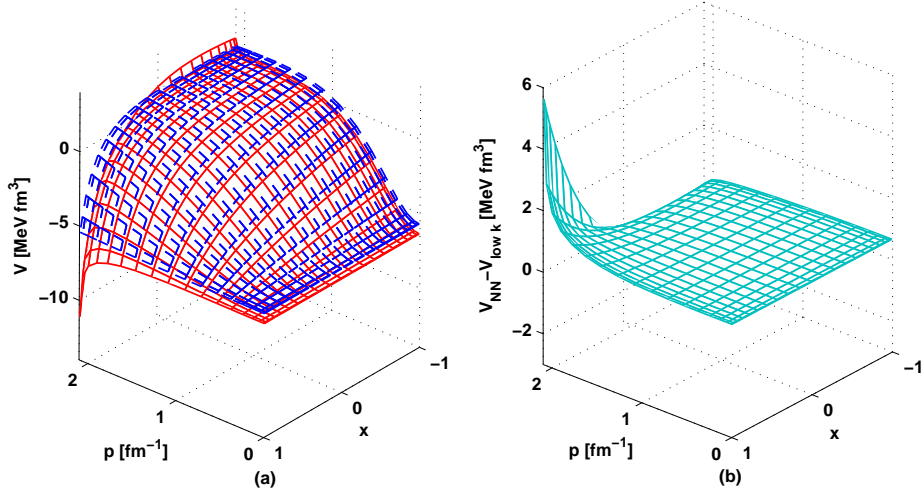}
\caption{\label{fig.vlowk} (a) The comparison of the low-momentum
effective potential $V_{low\,k}(p.p,x)$ (solid lines) with the
bare potential $V_{NN}(p,p,x)$ (dashed lines) and (b) differences
between them.}
\end{center}
\end{figure}

\begin{figure}
\begin{center}
\includegraphics*[width=10cm]{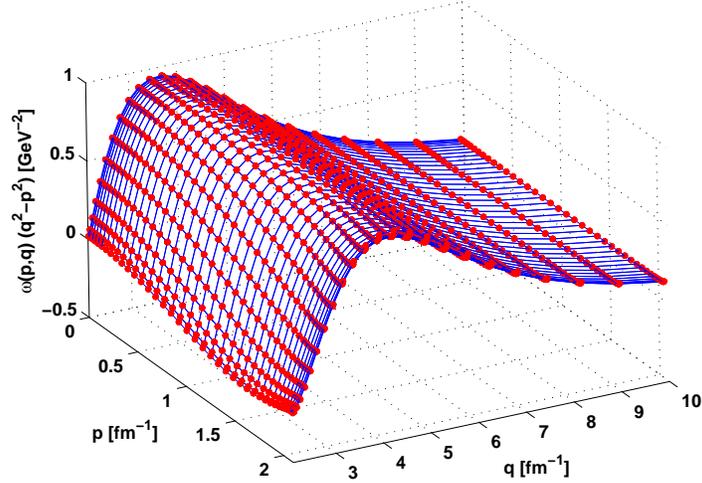}
\caption{\label{fig.Omega-Swave}The comparison of
$\omega(q,p)(q^{2}-p^{2})$ operator for the $S$-wave calculated in
the PW approach (dotes) and in the 3D approach (solid lines).}
\end{center}
\end{figure}

\begin{figure}
\begin{center}
\includegraphics*[width=10cm]{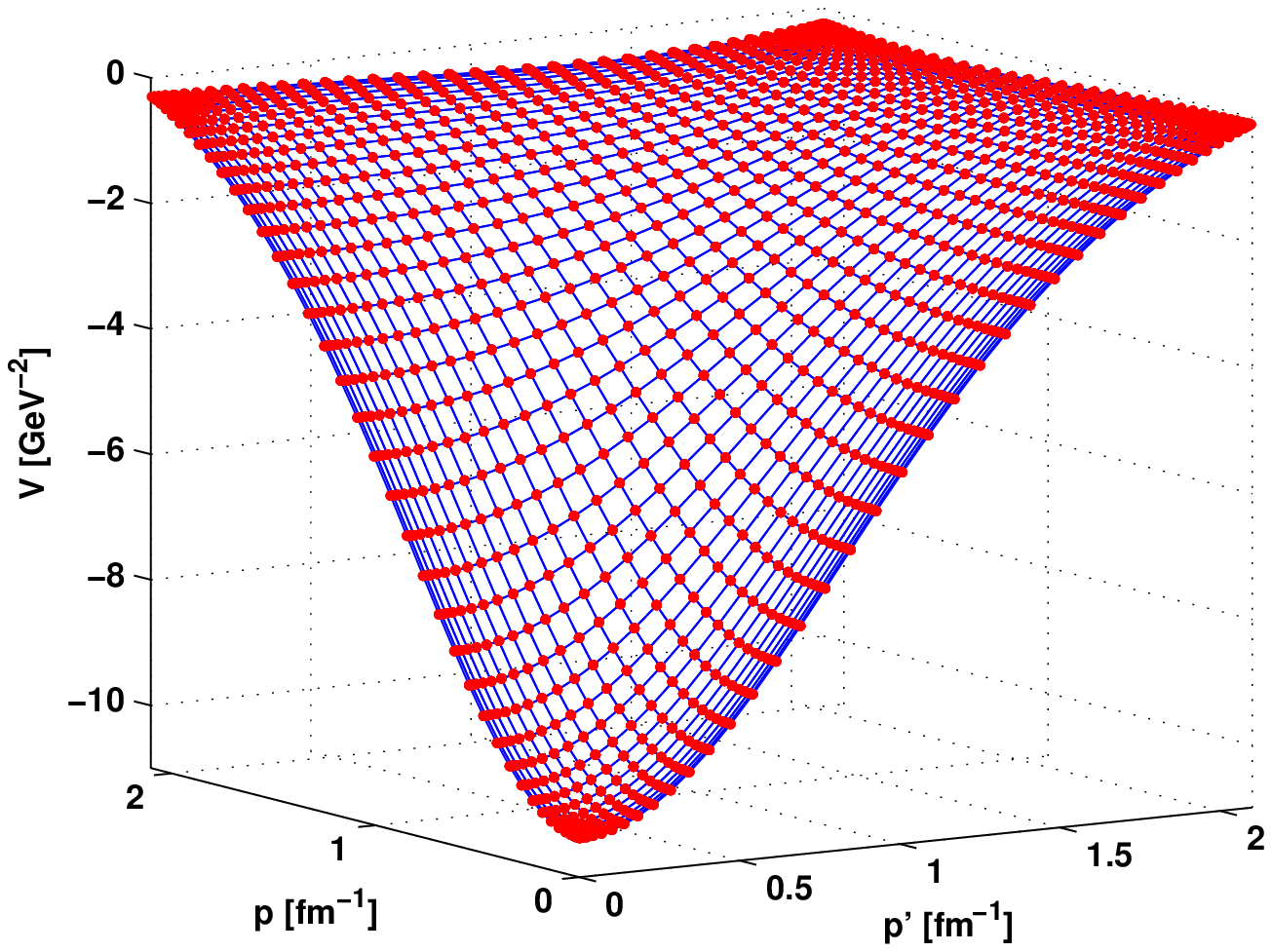}
\caption{\label{fig.Vlowk-Swave}The comparison of the low-momentum
effective potential $V_{_{low\,k}}(p',p)$ for the $S$-wave
calculated in the PW approach (dotes) and in the 3D approach
(solid lines).}
\end{center}
\end{figure}

We have shown in figure \ref{fig.cross section} that the
calculated total two-body cross section for the
$V_{low\,k}(p',p,x)$ and the $V_{NN}(p',p,x)$ match perfectly
well. Also we have calculated the $S$-wave phase shift for
$V_{low\,k}(p',p,x)$ and it has been compared with the obtained
result with the bare potential $V_{NN}(p',p,x)$ in figure
\ref{fig.phase shift}. For calculation of the phase shift we have
used the relation between the PW and 3D representations of the
on-shell $T$ matrix as follow:
\begin{eqnarray}
T_{l}(p,p,p^{2})=2\pi\int_{-1}^{1}dx\,\,\,T(p,p,x,p^{2})\,P_{l}(x).
\end{eqnarray}
As we expect the results are in good agreement with the high
accuracy. In figure (\ref{fig.vlowk}a) we have compared the
calculated low-momentum effective potential $V_{low\,k}(p,p,x)$
with the bare potential $V_{NN}(p,p,x)$, also In figure
(\ref{fig.vlowk}b) the difference between them is shown as a
function of the momentum variable $p$ and the angle variable $x$.
This comparison demonstrates that the difference between the two
potentials is approximately constant except for the momentum
values close to the cutoff, where the differences vary for
different angle values. Conceptually when we integrate on the
angles we can produce the same results as PW approach.

As a test of our calculations we can compare the obtained results
for the low-momentum effective potential and the $\omega$ operator
in both 3D and PW approaches. In the first step we directly
calculate the $\omega$ operator and the low-momentum effective
potential in PW approach. In the second step we obtain a certain
PW projection of the $\omega$ operator and the low-momentum
effective potential from their corresponding 3D representation by
the following relations:
\begin{eqnarray}
V^{l}_{low\,k}(p',p) &=&
2\pi\int_{-1}^{1}dx\,\,\,V_{low\,k}(p',p,x)\,P_{l}(x),
 \nonumber \\
\omega_{l}(q,p) &=&
2\pi\int_{-1}^{1}dx\,\,\,\omega(q,p,x)\,P_{l}(x).
\end{eqnarray}
The obtained results for the $S$-wave projection of the $\omega$
operator and the low-momentum effective potential in both of 3D
and PW approaches are given in figures ({\ref{fig.Omega-Swave})
and ({\ref{fig.Vlowk-Swave}) respectively. The agreement between
the two approaches is quiet satisfactory.

\section{Summary and Outlook}\label{sec:summary}
In this article the 3D formulation of the Lee-Suzuki and the RG
methods have been presented. The low-momentum effective
interaction $V_{low\,k}$ has been derived as a function of the
magnitude of momenta and the angle between them without using the
partial wave decomposition. The calculated two-body observables
from the low-momentum effective interaction and the bare
interaction have been shown. In addition a comparison of
calculated $V_{low\,k}$ from the PW and the 3D representation have
been demonstrated as a test for our calculations.

The advantage of our formulation in the 3D representation in
comparison with the PW representation is that we have calculated
the low-momentum effective interaction by considering all partial
waves automatically.

For the future investigations the low-momentum effective
interaction can be formulated by considering the spin and isospin
degrees of freedom in a realistic 3D approach. This formulation
based on the momentum helicity basis has been done and the
calculation of the realistic low-momentum effective interaction
with Bonn-B, AV18 and Chiral potentials is currently underway.
Considering the obtained 3D low-momentum effective interaction in
the few-body bound and scattering calculations is another major
task to be done.

\section*{Acknowledgments}

We would like to thank S. K. Bogner for fruitful discussion during
EFB19 conference about this work. This work was supported by the
research council of the University of Tehran.

\appendix

\section{RG method for $V_{low\,k}$ in the 3D momentum representation}\label{appendix:RG method}

For the two-body problem, the RG method derives the low-momentum
effective potential $V_{low\,k}$ by integrating out the model
dependent high momentum component of the different models of bare
potentials $V_{NN}$ by demanding that the low-momentum observables
calculated from $V_{NN}$ must be reproduced by $V_{low\,k}$ with
the same accuracy. We follow basically the same procedure which is
used by Bogner \emph{et al}. in derivation of the RG flow equation
\cite{Bogner-nt}. However instead of using the PW representation,
we have implemented the treatment of the RG in the 3D
representation. We start with a RG treatment of the scattering
problem in the 3D representation. The HOS two-body $T$ matrix in
3D representation by applying an arbitrary potential model
$V_{NN}$ to the scattering problem is given by the
Lippmann-Schwinger equation $(\hbar=m_{N}=1)$
\cite{Elster-FBS24}\footnotemark \footnotetext[1]{This quantity is
real in this appendix}:
\begin{eqnarray} \label{Lippmann-Schwinger-VNN}
T(\textbf{k}',\textbf{k},k^{2})=V_{NN}(\textbf{k}',\textbf{k})+\wp
\int_{0}^{\infty}\emph{d}p\,p^{2}\int\emph{d}\,\hat{\textbf{p}}
\frac{V_{NN}(\textbf{k}',\textbf{p})\,T(\textbf{p},\textbf{k},k^{2})}{k^{2}-p^{2}},
\end{eqnarray}
where $\wp$ denotes a principle value integration. Corresponding
to equation (\ref{eq.model space SE}) we can define a restricted
version of the equation (\ref{Lippmann-Schwinger-VNN}) by imposing
a cutoff $\Lambda$ and replacing the bare potential $V_{NN}$ by a
low-momentum effective potential $V_{low\,k}$ as:
\begin{eqnarray} \label{Lippmann-Schwinger-Vlowk}
T_{low\,k}(\textbf{k}',\textbf{k},k^{2})=V_{low\,k}(\textbf{k}',\textbf{k})+\wp
\int_{0}^{\Lambda}\emph{d}p\,p^{2}\int\emph{d}\,\hat{\textbf{p}}
\frac{V_{low\,k}(\textbf{k}',\textbf{p})\,T_{low\,k}(\textbf{p},\textbf{k},k^{2})}{k^{2}-p^{2}}. \nonumber \\
\end{eqnarray}
We demand that the calculated low-momentum HOS two-body \emph{T}
matrices from equations (\ref{Lippmann-Schwinger-VNN}) and
(\ref{Lippmann-Schwinger-Vlowk}) are identical, i.e.
$T(\textbf{k}',\textbf{k},k^{2})=T_{low\,k}(\textbf{k}',\textbf{k},k^{2})$
for $|\textbf{k}'|,\,|\textbf{k}|\leq \Lambda$. This condition
ensures that the $V_{low\,k}$ gives the same low-momentum two-body
observables as obtained by the bare potential $V_{NN}$. Imposing
the cutoff independence of the low-momentum HOS two-body $T$
matrix, i.e. $dT(\textbf{k}',\textbf{k},k^{2})/d\Lambda=0$ in
equation (\ref{Lippmann-Schwinger-Vlowk}), ensures that the
low-momentum observables will be independent of the scale
$\Lambda$ and will generate energy-independent potential
$V_{low\,k}$, therefore we obtain:
\begin{eqnarray}
\int_{0}^{\Lambda}\emph{d}\,p\,p^{2}\int\emph{d}\,\hat{\textbf{p}}\frac{dV_{lowk}(\textbf{k}',
\textbf{p})}{d\Lambda}\,\chi_{\textbf{k}}(\textbf{p})=\int
d\hat{\mathbf{p}}\frac{V_{lowk}(\textbf{k}',\Lambda\hat{\mathbf{p}})\,T(\Lambda\hat{\mathbf{p}},\textbf{k},k^{2})}{1-(k/\Lambda)^{2}},
\end{eqnarray}
where we have used the standing wave scattering states of the
effective theory $|\chi_{\textbf{k}}\rangle$, which can be written
as:
\begin{eqnarray}
|\chi_{\textbf{k}}\rangle=|\textbf{k}\rangle+\wp\int_{0}^{\Lambda}\emph{d}p\,p^{2}\int\emph{d}\,\hat{\textbf{p}}
\frac{T(\textbf{p},\textbf{k},k^{2})}{k^{2}-p^{2}}\,\,\,|\textbf{p}\rangle.
\end{eqnarray}
By using the completeness relation of the scattering states in the
model space and considering
$T(\Lambda\hat{\mathbf{p}},\textbf{k},k^{2})=\langle\Lambda\hat{\mathbf{p}}|V_{lowk}|\chi_{\textbf{k}}\rangle$,
we have obtained:
\begin{eqnarray}
\frac{d}{d\Lambda}\langle\,\textbf{k}'|V_{low\,k}|\textbf{p}'\rangle
&=& \Lambda^{2}\int
d\hat{\mathbf{p}}\,\,\langle\,\textbf{k}'|V_{low\,k}|\Lambda\hat{\mathbf{p}}\rangle
\nonumber \\ && \times \int_{0}^{\Lambda}dk\,k^{2}\int
d\,\hat{\textbf{k}}\frac{\langle\Lambda\hat{\mathbf{p}}|V_{low\,k}|\chi_{\textbf{k}}\rangle}
{\Lambda^{2}-k^{2}}\langle\tilde{\chi}_{\textbf{k}}|\textbf{p}'\rangle
\nonumber
\\ &=& \Lambda^{2}\int d\hat{\mathbf{p}}\langle\,\textbf{k}'|V_{low\,k}|\Lambda\hat{\mathbf{p}}\rangle \int_{0}^{\Lambda}dp''p''^{2}\int
d\,\hat{\textbf{p}}''\langle\Lambda\hat{\mathbf{p}}|V_{low\,k}|\textbf{p}''\rangle
\nonumber \\ && \times \int_{0}^{\Lambda}dk\,k^{2}\int
d\,\hat{\textbf{k}}\frac{\langle\textbf{p}''|\chi_{\textbf{k}}|\rangle\langle\tilde{\chi}_{\textbf{k}}|\textbf{p}'\rangle}
{\Lambda^{2}-k^{2}}\nonumber
\\ &=& \Lambda^{2}\int d\,\hat{\mathbf{p}}\langle\textbf{k}'|V_{low\,k}|\Lambda\hat{\mathbf{p}}\rangle \nonumber \\ && \times \int_{0}^{\Lambda}dp''p''^{2}\int
d\,\hat{\textbf{p}}''\langle\Lambda\hat{\mathbf{p}}|V_{low\,k}|\textbf{p}''\rangle
G(\textbf{p}'',\textbf{p}',\Lambda^{2}),
\end{eqnarray}
where $\langle\tilde{\chi}_{\textbf{k}}|$ is the bi-orthogonal
complement in completeness relation in the model space which
satisfies
$\langle\tilde{\chi}_{\textbf{k}'}|\chi_{\textbf{k}}\rangle=\delta(\textbf{k}'-\textbf{k})$
and $G$ denotes the interacting Green's function in the model
space. With writing the $G$ in terms of the two-body $T$ matrix we
obtain the RG equation in the 3D momentum representation as:
\begin{eqnarray}
\frac{d}{d\Lambda}V_{low\,k}(\textbf{k}',\textbf{k})=\int
d\hat{\mathbf{p}}
\frac{V_{low\,k}(\textbf{k}',\Lambda\hat{\mathbf{p}})\,
T(\Lambda\hat{\mathbf{p}},\textbf{k},\Lambda^{2})}{1-(k/\Lambda)^{2}}.
\end{eqnarray}
For numerical calculations we can choose the suitable coordinate
system where the vector $\textbf{k}'$ is along the $z$ axis and
the vector \textbf{k} is in $x-z$ plane. By this consideration we
can rewrite the above equation as:
\begin{eqnarray}
\frac{d}{d\Lambda}V_{low\,k}(k',k,x)=\int_{-1}^{1}
dx'\int_{0}^{2\pi}d\varphi\frac{V_{low\,k}(k',\Lambda,x')\,
T(\Lambda,k,y,\Lambda^{2})}{1-(k/\Lambda)^{2}},
\end{eqnarray}
where:
\begin{eqnarray}
 x&=&\hat{\textbf{k}}'\cdot\hat{\textbf{k}}, \nonumber \\
x'&=&\hat{\mathbf{k}}'\cdot\hat{\textbf{p}}, \nonumber \\
y&=&\hat{\textbf{k}}\cdot\hat{\mathbf{p}}=xx'+\sqrt{1-x^{2}}\sqrt{1-x'^{2}}\cos\varphi,
\nonumber \\ k&=&|\,\textbf{k}|, \nonumber \\
k'&=&|\,\textbf{k}'|.
\end{eqnarray}
We introduce:
\begin{eqnarray}
\hat{T}(k',k,x',x)=\int_{0}^{2\pi}d\varphi\,T(k',k,x'x+\sqrt{1-x'^{2}}\sqrt{1-x^{2}}\cos\varphi).
\end{eqnarray}
And finally the RG equation in the 3D representation can be
obtained:
\begin{eqnarray}
\frac{d}{d\Lambda}V_{low\,k}(k',k,x)=\int_{-1}^{1}
dx'\frac{V_{low\,k}(k',\Lambda,x')\,
\hat{T}(\Lambda,k,x',x,\Lambda^{2})}{1-(k/\Lambda)^{2}}.
\end{eqnarray}
The low-momentum effective interaction $V_{low\,k}(k',k,x)$ can be
calculated by numerically integrating the RG equation on the
Gaussian momentum and angle grid points. Similar to the PW
representation of this equation, $V_{NN}(k',k,x)$ can be used as a
large-cutoff initial condition to calculate $V_{low\,k}(k',k,x)$
numerically.


\begin{thebibliography}{}


\bibitem{Bogner-PR386} S. K. Bogner, T. T. S. Kuo and A. Schwenk, Phys. Rep. 386 (2003)
1.

\bibitem{Bogner-PLB576} S. K. Bogner, T. T. S. Kuo, A. Schwenk, D.R. Entem and R. Machleit, Phys. Let. B576 (2003)
265.

\bibitem{Epelbaum-PLB439} E. Epelbaum, W. Gl\"{o}ckle, U.G Mei{\ss}ner, Phys. Let. B439 (1998)
1.

\bibitem{Bogner-nt} S. K. Bogner,  A. Schwenk,  T. T. S. Kuo and G. E. Brown,
nucl-th/0111042.

\bibitem{Fujii-PRC70} S. Fujii, E. Epelbaum, H. Kamada, R. Okamoto, K. Suzuki and W. Glockle, Phys. Rev. C70 (2004)
024003.

\bibitem{Birse-PLB464} M. C. Birse, J. A. McGovern and K. G. Richardson, Phys. Lett. B464 (1999)
169.

\bibitem{Suzuki-PTP68} K. Suzuki, Prog. Theor. Phys. 68 (1982) 246.

\bibitem{Lee-PLB91} S. Y. Lee and K. Suzuki, Phys. Lett. B91 (1980)
173.

\bibitem{Suzuki-PTP92} K. Suzuki and R. Okamoto, Prog. Theor. Phys. 92 (1994)
1045.

\bibitem{Suzuki-PTP64} K. Suzuki and S.Y. Lee, Prog. Theor. Phys. 64 (1980)
2091.

\bibitem{Suzuki-PTP70} K. Suzuki and R. Okamotp, Prog. Theor. Phys. 70 (1983)
439.

\bibitem{Elster-FBS24} Ch. Elster, J. H. Thomas and W. Gl\"{o}ckle, Few Body Systems 24 (1998)
55.

\bibitem{Elster-FBS27} Ch. Elster, W. Schadow, A. Nogga and W. Gl\"{o}ckle, Few Body Systems 27 (1999)
83.

\bibitem{Schadow-FBS28} W. Schadow, Ch. Elster and W. Gl\"{o}ckle, Few Body Systems 28 (2000)
15.

\bibitem{Fachruddin-PRC62} I. Fachruddin, Ch. Elster and W. Gl\"{o}ckle, Phys. Rev. C 62 (2000) 044002.

\bibitem{Fachruddin-PRC63} I. Fachruddin, Ch. Elster and W. Gl\"{o}ckle, Phys. Rev. C 63 (2001)
054003.

\bibitem{Liu-FBS33} H. Liu, Ch. Elster and W. Gl\"{o}ckle, Few Body Systems 33 (2003) 241.

\bibitem{Fachruddin-MPLA18} I. Fachruddin, Ch. Elster and W. Gl\"{o}ckle, Mod. Phys. Lett. A18 (2003)
452.

\bibitem{Fachruddin-PRC68} I. Fachruddin, Ch. Elster and W. Gl\"{o}ckle, Phys. Rev. C68 (2003)
054003.

\bibitem{Fachruddin-PRC69} I. Fachruddin, W. Gl\"{o}ckle, Ch. Elster and A. Nogga, Phys. Rev. C69 (2004)
064002.

\bibitem{Liu-PRC72} H. Liu, Ch. Elster and W. Gl\"{o}ckle, Phys. Rev. C72 (2005)
054003.

\bibitem{Lin-PLB660} T. Lin, Ch. Elster, W. N. Polyzou, and W. Gl\"{o}ckle, Phys. Let. B660 (2008) 345.

\bibitem{Hadizadeh-FBS40} M. R. Hadizadeh and S. Bayegan, Few Body Systems 40 (2007)
171.

\bibitem{Hadizadeh-EPJA} M. R. Hadizadeh and S. Bayegan, Eur. Phys. J. A36 (2008)
201.

\bibitem{Bayegan-PRC77} S. Bayegan, M. R. Hadizadeh and M. Harzchi, Phys. Rev. C77 (2008)
064005.

\bibitem{Bayegan-EFB20} S. Bayegan, M. R. Hadizadeh and M. Harzchi, to appear in Few Body Systems,
arXiv:0711.4036.

\bibitem{Bayegan-PTP} S. Bayegan, M. R. Hadizadeh and W. Gl\"{o}ckle, Submitted to Prog. of Theor. Phys.
arXiv:0806.1520.

\bibitem{Bayegan-p} S. Bayegan, M. Harzchi and M. R. Hadizadeh, \emph{in
preparation}.

\bibitem{Jennings-EL72} B. K. Jennings, Europhys. Lett 72 (2005) 211.

\bibitem{Bogner-PLB500} S. K. Bogner and T.T.S. Kuo, Phys. Lett. B500 (2001)
279.

\bibitem{Malfliet-NPA127} R. A. Malfliet and J. A. Tjon, Nucl. Phys. A127 (1969)
161.

\bibitem{59} D. H\"{u}ber, H. Witala, A. Nogga, W. Gl\"{o}eckle and H. Kamada, Few Body Systems 22 (1997)
107.

\bibitem{60} the routines called ZGESV from
{http://netlib.org/lapack/double/}

\end{thebibliography}
\end{document}